# Dimensions: A Competitor to Scopus and the Web of Science?

Mike Thelwall, University of Wolverhampton, UK.

Dimensions is a partly free scholarly database launched by Digital Science in January 2018. Dimensions includes journal articles and citation counts, making it a potential new source of impact data. This article explores the value of Dimensions from an impact assessment perspective with an examination of Food Science research 2008-2018 and a random sample of 10,000 Scopus articles from 2012. The results include high correlations between citation counts from Scopus and Dimensions (0.96 by narrow field in 2012) as well as similar average counts. Almost all Scopus articles with DOIs were found in Dimensions (97% in 2012). Thus, the scholarly database component of Dimensions seems to be a plausible alternative to Scopus and the Web of Science for general citation analyses and for citation data in support of some types of research evaluations.

## 1. Introduction

Citation counts are used by researchers and research managers to help evaluate the quality or impact of published research, particularly when it is impractical to employ peer judgements or a second opinion is needed. In the early years of citation analysis there was a single pre-eminent data source for citation counts, Eugene Garfield's Science Citation Index (Garfield, 1964) but today Scopus has become a viable alternative (Archambault, Campbell, Gingras, & Larivière, 2009; Falagas, Pitsouni, Malietzis, & Pappas, 2008). There are also free online citation indexes, such as Google Scholar (Halevi, Moed, & Bar-Ilan, 2017; Martin-Martin, Orduna-Malea, Harzing, & López-Cózar, 2017; Prins, Costas, van Leeuwen, & Wouters, 2016) and Microsoft Academic (Harzing & Alakangas, 2017; Hug, Ochsner, & Brändle, 2017; Sinha, Shen, Song, Ma, Eide, Hsu, & Wang, 2015; Thelwall, 2018). The existence of alternatives has three main benefits for research evaluators. First, the free alternatives may reduce the cost of evaluations and make informal impact self-evaluations possible for many researchers that would not pay to access data. Second, all citation indexes are imperfect and the availability of alternatives allows data from one to be cross-checked against the alternatives. Third, each citation index may have coverage advantages or capabilities that make it a better fit for a given impact evaluation task.

In January 2018, Digital Science launched Dimensions, a new online scholarly platform for publications, grants, clinical trials and patents, giving free partial online access (Adams, Draux, Jones, Osipov, Porter, & Szomszor, 2018). The platform replaced a previous grant analysis tool, also called Dimensions, to support "portfolio analysis and planning for science funders"[1]. This article focuses on the publication component of Dimensions. Publications in Dimensions are categorised as articles (75,698,402 on 19 February 2018), chapters (9,525,334), proceedings (4,975,857), monographs (328,484) and preprints (19,734). The relatively small numbers of preprints and lack of other sources suggests that the Dimensions data is predominantly from publishers. The preprints originate from bioRxiv, and, according to its founder Christian Herzog (personal communication) and (private) developer FAQ, it plans to index more preprint archives and some institutional repositories. In contrast, Microsoft Academic and Google Scholar also index web content from crawlers (Halevi, Moed, & Bar-Ilan, 2017; Harzing & Alakangas, 2017). This presumably means that the citation counts in Dimensions are lower than those of Microsoft Academic and Google Scholar but has the important advantage that its data is less easy to spam. Generating fake papers with self-citations and posting them to academic domains is an effective way to spam indexes that look for academic content online (Delgado López-Cózar, Robinson-García, & Torres-Salinas, 2014). Whilst there are also low quality academic journals (Gutierrez, Beall, & Forero, 2015) and unethical practices in peer reviewed publications (Chorus, 2015), these can be policed by the academic community (e.g., the Committee on Publication Ethics (COPE) claims 12,000 members: publicationethics.org/about) and are not as powerful as unlimited self-publishing to academic domains. Dimensions might also contain fewer data processing errors a result of avoiding web data. It therefore apparently fulfils a unique niche as a large scale partly free citation index that is protected against spam.

---
[1] https://web.archive.org/web/20170820121951/http://www.uberresearch.com:80/dimensions-for-funders/

Given the potential value of Dimensions for research evaluations, it is important to assess its key properties to decide whether it contains enough data to be useful and whether its citation counts are plausible.

## 2. Research questions

The aim of this study is to give insights into Dimensions rather than to provide comprehensive information. As a young service, it may evolve soon, undermining the value of a detailed empirical analysis. Dimensions is compared to Scopus but not the Web of Science since Scopus has consistently been found to have greater overall coverage of academic journals (Mongeon & Paul-Hus, 2016; Waltman, 2016) and so represents best practice in terms of comprehensiveness. The following exploratory research questions drive the study.

1. How comprehensive is the coverage of Scopus journal articles in Dimensions?
2. Are the average citation counts for journal articles in Dimensions comparable to those of Scopus?
3. Are Dimensions citation counts for journal articles interchangeable with those of Scopus, in the sense of having a very high correlation with them?

## 3. Methods

The Scopus narrow category Food Science was chosen for an exploratory analysis. This is an average category from the perspective of online citation counts, with the median average (geometric mean) Microsoft Academic citation count of all Scopus narrow fields in a recent study (Thelwall, 2018). The years 2008-2017 were selected to allow an analysis of changes over time in the results and the data was collected in February 2018 from Scopus using its Applications Programming Interface (API). Citation counts and DOIs were extracted from the Scopus records. Articles without DOIs were discarded since these could not be easily matched accurately. Only documents recorded in Scopus as (standard) journal articles were used, excluding books, conference papers, reviews and editorials, for example.

In February 2018 Dimensions was queried by DOI for all journal articles with DOIs returned by the Scopus queries (n=84691) and the Dimensions citation counts were recorded.

For RQ1, the coverage of Dimensions was compared against Scopus as a benchmark by calculating for each year the percentage of Scopus articles with DOIs that were also in Dimensions with a DOI. This is a one-way comparison since Dimensions may cover articles that are not in Scopus but should nevertheless give broad insights into whether Dimensions has substantial coverage of science.

To check whether Food Science is an unusual case, a random sample of 10000 articles with DOIs from Scopus in 2012 was also checked for matching records in Dimensions. This sample was selected using a random number generator from a list of the most recent 5000 (a system limitation) articles in all 326 non-empty Scopus narrow fields. Recycled Scopus data that had originally been collected 26 August 2017 for another paper was used for this.

For RQ2, the geometric mean citation counts for Dimensions each year were compared against those of Scopus. For this calculation, two different comparisons were made. For the first, articles not found in Dimensions were excluded. For the second, articles not found in Dimensions were included and given a citation count of 0. The geometric mean is a better measure of central tendency than the arithmetic mean because citation data is highly skewed (Fairclough & Thelwall, 2015).

For RQ3, citation counts from Scopus and Dimensions were compared for each year using Spearman correlations, as appropriate for skewed data. A high correlation suggests that the two may be interchangeable in practice for impact calculations. The correlations were calculated for data with and without articles not found in Dimensions, as for RQ2.

Although not directly addressing the research questions, Altmetric Scores and RCR (Relative Citation Ratio) values were also extracted from the data to provide additional context about Dimensions. Altmetric Scores provided by Dimensions are derived from Altmetric.com (Adie & Roe, 2013), and are weighted sums of all transparent scores collected by Altmetric, including citations from blogs, Twitter and Facebook but not Mendeley. RCR uses the co-citation network of an article to normalise its citation count (Hutchins, Yuan, Anderson, & Santangelo, 2016). This is more sensitive to the field of an article than the field of the journal publishing the article. This indicator has been criticised for a lack of transparency and technical problems with the calculation, such as with the method used to estimate the publication field of an article (Janssens, Goodman, Powell, & Gwinn, 2017). Dimensions also reports a Field Citation Ratio, which is field normalised citation score (for background theory, see: Waltman, van Eck, van Leeuwen, Visser, & van Raan, 2011). This was not included because it is not currently part of the Dimensions API.

## 4. Results and discussion

Almost all older Scopus Food Science articles with DOIs could be found in Dimensions via DOI searches for all years (Figure 1) except the partial current year (2018). Presumably Dimensions is slower to index some journals than Scopus and will have similarly high coverage of articles from 2018 shortly after the end of this year. About half of the articles have an RCR score and, for recent articles, about a fifth have an Altmetric Score. Uncited articles and articles under 2 years old are not given an RCR score, according to the Dimensions documentation (23 February 2018 via the help button on a Dimensions web search results) but there were also older cited articles without an RCR value such as one from 2015 with 17 citations (Dimensions ID: pub.1049677485).

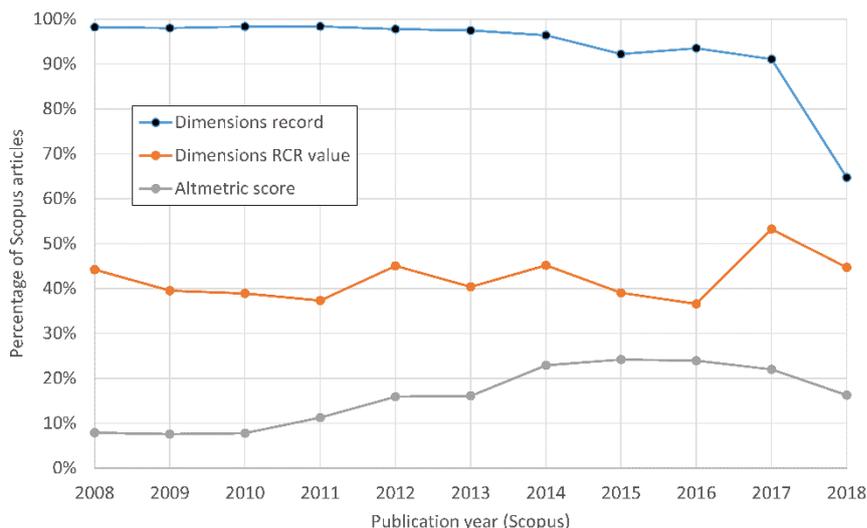

Figure 1. Dimensions coverage of Scopus Food Science articles with DOIs 2008-2018 (February) as of February 2018.

Average (geometric mean) citation counts for articles in Dimensions are about the same as Scopus citation counts (Figure 2). The Scopus line excludes the small percentage (for most years) of articles not found in Dimensions so the results are directly comparable. If these were included then most Scopus averages would be 3% lower.

The Altmetric scores and RCR values are not directly comparable to citation counts. RCR scores are normalised and therefore should not increase or decrease over time, as reflected in the flat line close to 1 (the world average value). Altmetric averages are stable over time, reflecting articles being discussed most in social media at the time when they are published, so there is little increase in the longer term. The higher Altmetric averages for older articles are due to the small percentage of articles with a score being important enough to be mentioned long after they were published (Altmetric.com started collecting social media data in 2011).

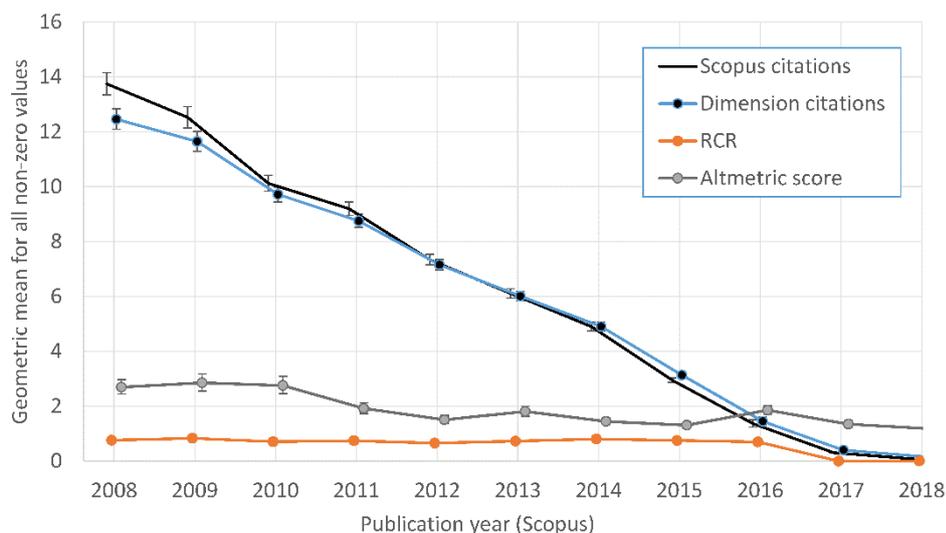

Figure 2. Average (geometric mean) scores for Scopus Food Science articles 2008-2018 (February) as of February 2018. For each source, all articles without a score are excluded for the calculations. For Scopus, articles not found in Dimensions are excluded from the calculations. Error bars show 95% confidence intervals.

There is a very high correlation between the citation counts from Scopus and Dimensions for articles indexed in both, especially in the longer term (Figure 3). The lower correlations for more recent years are presumably due to (a) the smaller numbers involved so that the presence or absence of individual citations can make a bigger difference and (b) the shorter time to accumulate citations making differences in coverage for citing sources have a proportionally larger effect on the total citation counts.

The high correlations between RCR values and Scopus citation counts indicate that the RCR calculations do not have a substantial effect in terms of estimating the impact of Food Science articles (Figure 3). This field might be relatively homogenous in terms of citation patterns, for example. The low correlations between Scopus citation counts and Altmetric.com Scores is in line with previous studies with altmetric data (Costas, Zahedi, & Wouters, 2015; Thelwall, Haustein, Larivière, & Sugimoto, 2013). Whilst Mendeley reader counts have moderate or strong correlations with Scopus citation counts (Haustein, Larivière, Thelwall, Amyot, & Peters, 2014; Thelwall, 2017) they are not included in the main Altmetric Score because they are not transparent. Altmetric.com reports Mendeley reader counts (Adie, 2013), but separately from its flagship Altmetric Score because it is not possible to check which Mendeley users were the source of the Mendeley reader counts.

The low correlation between RCR values and Scopus citation counts for 2017 and 2018 seems to be a system error. Dimensions reported RCR scores of 0 for many articles from these years that were uncited. These articles should not have been given an RCR value according to the Dimensions documentation since they were uncited and younger than two years old.

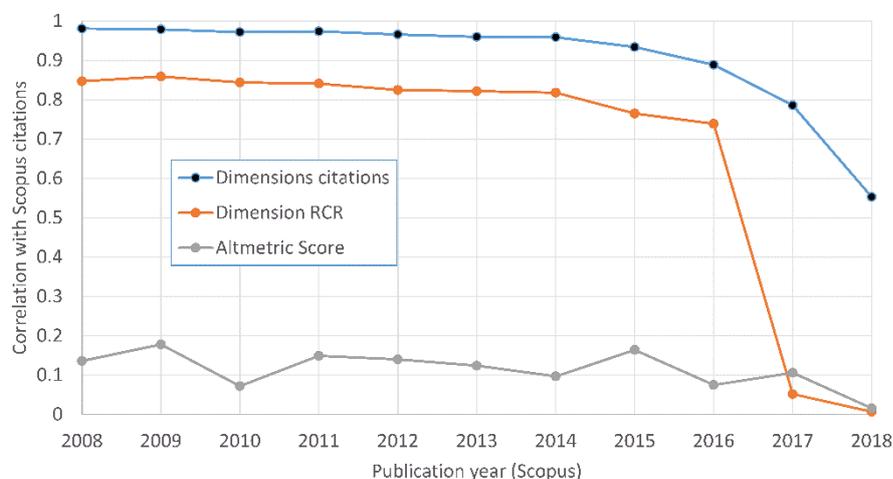

Figure 3. Spearman correlations with Scopus citations for Scopus Food Science articles 2008-2018 (February) as of February 2018. For each source, missing articles are excluded from the calculation.

Food Science articles from 2012 were investigated to obtain further insights (Figure 4). Publications with the largest citation count difference between Dimensions and Scopus were investigated for evidence of the causes (three in each direction). Google Scholar was also consulted as an independent evidence source. The following results were found.

- *Relationships between rumination time, metabolic conditions, and health status in dairy cows during the transition period* (Dimensions: 0 citations; Scopus: 35 citations; Google Scholar: 65 citations). This article has an erratum online (J Anim Sci. 2013

Mar;91(3):1522). Whilst Dimensions points to the correct source and erratum, attributing no citations to the latter, the erratum may have confused its citation counting algorithm.

- *Plant physiological responses to UV-B radiation* (D: 11; S: 42; GS: 67). At the time of checking (11 March 2018) the publisher website Ejfa.Info was not working and was for sale so it is possible that publisher self-citations had been lost by Dimensions due to an inability to fully index the site. In contrast, Scopus and Google Scholar would have had longer to index citations from this publisher.
- *Potato consumption and cardiovascular disease risk factors among Iranian population* (D: 54; S: 84; GS: 107). The citations in Scopus but not Dimensions were from Der Pharma Chemica (3), Der Pharmacia Lettre (9), Journal of Chemical and Pharmaceutical Research (4), Journal of Mazandaran University of Medical Sciences (2), Journal of Babol University of Medical Sciences (2), Journal of Pharmaceutical Sciences and Research (2), and nine other journals. A search for the journal names found no results in Dimensions, suggesting that it had not indexed this set of journals. The remaining two citations in Scopus but not Dimensions were from two articles indexed twice by Scopus (once incorrectly in each case). Thus, the main cause of the difference in this case was coverage by Scopus of journals not indexed by Dimensions in areas related to the article topic.
- *Total polyphenols, total flavonoid contents, and antioxidant activity of Korean natural and medicinal plants* (D: 62; S: 35; GS: 110). The citations in Dimensions but not Scopus were mainly from journals not indexed by Scopus, including Asian Journal of Beauty and Cosmetology (3), Journal of Life Science (3), Journal of the East Asian Society of Dietary Life (2), Korean Journal of Medicinal Crop Science (3), The Korean Journal of Food and Nutrition (4) and 14 other journals. Scopus had found only 3 of the 8 citations from Korean Journal of Food Preservation. Scopus started indexing this journal in 2017, so did not find the earlier citations. Overall, however, the main reason for the higher Dimensions citation count seems to be its more substantial coverage of related journals.
- *Vegetable breeding in Africa: constraints, complexity and contributions toward achieving food and nutritional security* (D: 30; S: 5; GS: 45). There were 25 additional Scopus citations to a second copy of this article, with Scopus recording the journal name as "Food Secur." for this version (without a DOI) and "Food Security" (with a DOI) for the main version (i.e., tied to a record from the journal). Thus, Scopus has not recognised the journal name abbreviation in this case, an algorithmic processing error.
- *Acaricide treatment affects viral dynamics in varroa destructor-infested honey bee colonies via both host physiology and mite control* (D: 57; S: 34; GS: 90). Scopus indexed two versions of this article and split the citations between them. The second version was an erratum to the article (a set of corrected graphs). The problem was therefore with incorrect merging of different entry points to the same article.

Summarising the above, whilst the citation counts are similar overall between Scopus and Dimensions, there can be large differences for a small number of articles due to indexing errors in either database or journal coverage differences that affect the articles' topics.

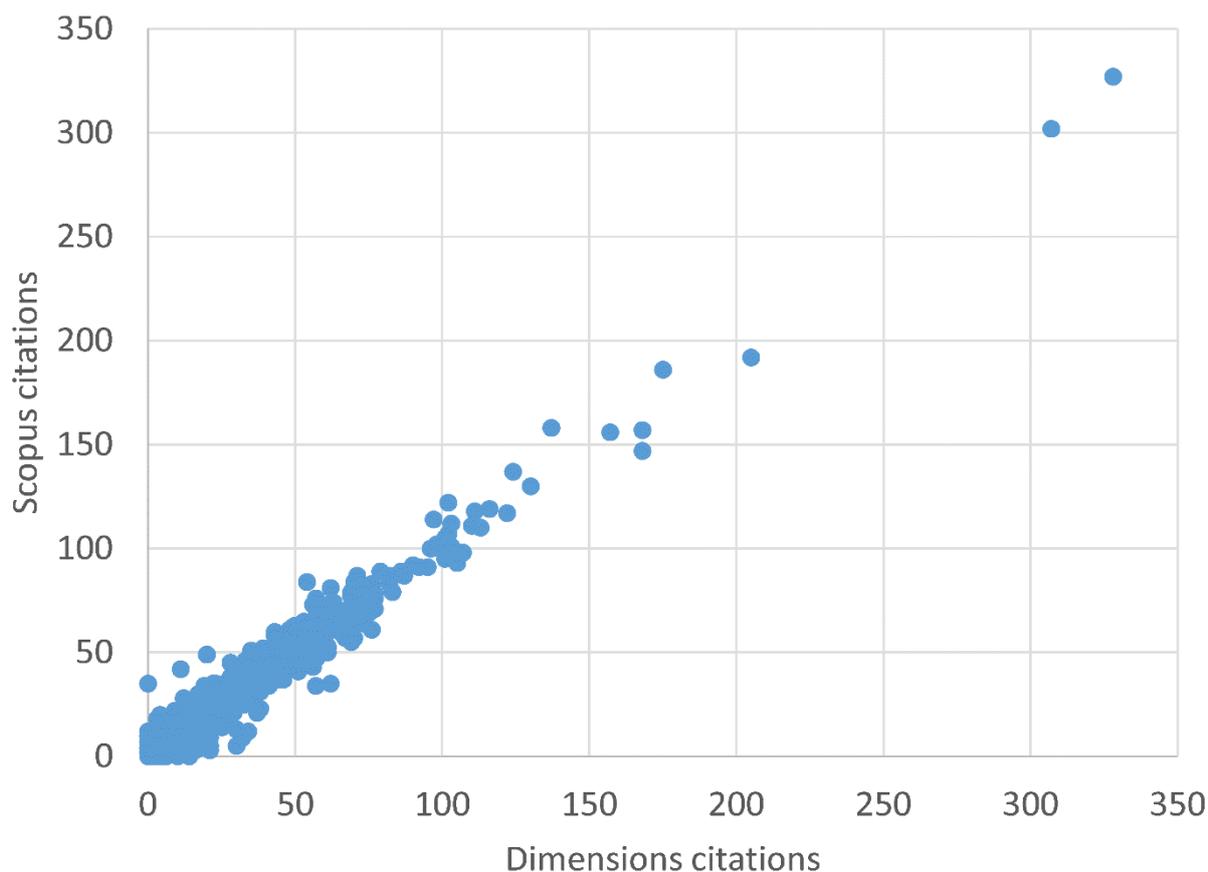

Figure 4. Scopus citations against Dimensions citations for Scopus Food Science articles with DOIs from 2012, as of February 2018 (n=7589). Articles not in Dimensions are excluded.

For the random sample of 10,000 articles from Scopus in 2012, almost all (9711; 97%) were found in Dimensions with a DOI search. Correlations can be misleadingly high for multidisciplinary article sets that combine high and low citation specialisms so Spearman correlations were calculated separately for each field and the median taken, giving 0.958. Taken together, these confirm that Scopus and Dimensions are interchangeable as data sources, in terms of coverage and citation counts. The median (across fields) geometric mean citation count for Scopus (August 2017) was 5.99, which was slightly lower than the 6.32 for Dimensions (February 2018). The difference is presumably due to the extra half year for citations to accrue. There was a median of 57 articles per field in this test.

Dimensions RCR values were available for 7450 of the 10000 random Scopus articles from 2012, with a median Spearman correlation with Scopus citation counts of 0.838 across the 240 fields with at least 2 randomly selected articles. Altmetric scores were available for 4273 of the 10000 random Scopus articles from 2012, with a median Spearman correlation with Scopus citation counts of 0.262 across the 281 fields with at least 2 randomly selected articles.

Individual correlations, sample sizes and geometric means for each field for the 10,000 articles from the 2012 dataset are available in Figshare (doi:10.6084/m9.figshare.5970952).

## 5. Conclusions

This article has assessed one aspect of the new scholarly database Dimensions, its journal articles. It has not investigated monographs, papers, chapters, preprints. It has also not investigated the novel features of the site, such as its grants, patents and clinical trials data, as well as its analytical tools.

Combining the results for one field 2008-2018 and those from the 2012 random sample from all fields, the results suggest that the coverage and citation counts of Dimensions are comparable to those of Scopus. Thus, Dimensions seems to be essentially interchangeable with Scopus in terms of coverage and citation counts. Large citation count differences for a few individual articles can nevertheless occur due to indexing errors or differences in journal coverage related to the article topics. This article has not tested whether Dimensions covers articles that are not in Scopus. Presumably it does not cover many of these because its citation counts seem to be slightly lower than those of Scopus.

Unlike Google Scholar and Microsoft Academic, Dimensions seems to mainly index peer reviewed articles, except for the preprint server bioRxiv (and other scholarly repositories in the future) and presumably also a small percentage of unrefereed journal articles, scholarly monographs and book chapters. It may therefore be possible to use it for some formal research evaluation purposes unless or until it substantially extends its coverage of unrefereed content. If this unrefereed content expansion occurs then it would need to report refereed citation counts separately to maximise its value for formal research evaluations.

Since Dimensions is currently indirectly spammable through preprint servers (e.g., by uploading batches of low quality content), in its current form it should not be used for bibliometrics-driven research evaluations, such as the Italian system (Franceschet & Costantini, 2011) because of the direct financial incentive to game citation counts (see also a related argument for altmetrics: Wouters & Costas, 2012). For other evaluation systems, such as the UK REF, where citations play a relatively minor role in supporting peer review (Wilsdon, Allen, Belfiore, Campbell, Curry, Hill, & Jones, 2016), the indirect financial incentive may not outweigh the reputational drawback from posting low quality content. An honesty clause may still be needed to guard against justifiable manipulations (e.g., Delgado López-Cózar, Robinson-García, & Torres-Salinas, 2014). Overall, however, the results suggest that Dimensions is a competitor to the Web of Science and Scopus for non-evaluative citation analyses and for supporting some types of formal research evaluations.